\documentclass[english]{lni}

\usepackage{booktabs}

\usepackage[]{blindtext}
\usepackage{amsmath}
\usepackage{tcolorbox}
\usepackage{multicol}

\begin{document}
\title[]{Experimental Evaluation of Dynamic Topic Modeling Algorithms}
 \author[1]{Ngozichukwuka Onah}{ngozichukwuka.onah@tu-ilmenau.de}{0009-0004-1668-0925}
 \author[2]{Nadine Steinmetz}{nadine.steinmetz@fh-erfurt.de}{0000-0003-3601-7579}
 \author[1]{Hani Al-Sayeh}{hani-bassam.al-sayeh@tu-ilmenau.de}{0000-0002-4381-6865}
 \author[1]{Kai-Uwe Sattler}{kus@tu-ilmenau.de}{0000-0003-1608-7721}%
 \affil[1]{TU Ilmenau, DBIS, Helmholtzplatz 5, 98693 Ilmenau}
 \affil[2]{University of Applied Sciences Erfurt, Altonaer Str. 25, 99085 Erfurt}

\maketitle

\begin{abstract}
The amount of text generated daily on social media is becoming gigantic and analyzing this text is useful for many purposes including 
exploring marketing trends, public opinion, and sentiment analysis.
To fully understand what lies beneath a huge amount of text, 
scientific researchers need dependable and effective computing techniques from self-powered topic models like Latent Dirichlet Allocation Sequence, 
Hierarchical Dirichlet Process, 
Bidirectional Encoder Representations from Transformers Topic, 
and Top2Vec. 
Nevertheless, there are currently relatively few thorough quantitative comparisons 
between these models, particularly 
the investigation of how they behave when modeling topics in an event-driven context and identifying topics that change over time. 
In this study, we compare these models with existing evaluation metrics and propose an assessment metric that documents how the topics change in an event-driven setting. 
Our experiments show that each model's performance depends on the goal. 
For example, while LDASequence requires more hardware resources, BERTopic is preferred for excellent topic quality, Top2Vec topics evolution is high,   
and HDP execution times are low.
\end{abstract}



\section{Introduction}
In recent times, there has been a transformation in the utilization of technologies and heightened human communication via social media platforms \cite{abdelrazek2023, Joubert2019}. 
On social media platforms (e.g. Facebook, Instagram, and Twitter ), people share perspectives (e.g. comments/texts, audios, videos) on a wide range of topics including politics, 
economy, natural disasters, and 
infectious diseases like coronavirus \cite{angelov2020, ghaisani2017}.  
 Obtaining insight into opinions, views, and thoughts from comments is crucial, particularly when it comes to situations or instances that demand immediate action, such as viral illnesses. Thus, making notes of the key terms from events as they unfold in real-time will help with instant decision-making as well as raising one's knowledge of current events. 

The process of finding the most important terms (called \textit{topics}) associated with documents is known as \textit{topic modeling}. It provides an outline of a document's contents, thereby summarizing the main idea presented in the document \cite{egger2022a}. 
Real-world events like corona cases, lock-downs, border closures, vaccine campaigns, and vaccinations took place at specific time segments  \cite{lau2020, le2020, vijayan2021}. 
 Finding topics that are time-dependent becomes important in a situation where rapid decision making is very important. \textit{Dynamic Topic Models (DTMs)} are topic models that are applied in relation to time \cite{blei2006}. 

Dynamic Topic Modeling in this context can be applied by considering texts (e.g. comments) as a document and discovering events (e.g.lock-downs) as topics. DTMs help to identify how topics change over time in sequentially structured documents \cite{blei2006, dieng2019, bhadury2016}. 
There are varieties of algorithms from both the old and state-of-the-art algorithms that are used to build both static and dynamic topic models and they have been the subject of research, performance assessments, and analysis \cite{egger2022a, Pavithra2024a, montero2023, james2023, Pavithra2024b}. 
Yet, in an event-driven environment (e.g. real-time system for monitoring opinions of twitter users on covid19) where texts are studied to gain an opinion of recent happenings on daily, monthly, or quarterly bases, a thorough examination and detailed comparison between algorithms that can be applied in dynamic topic modeling is still missing, especially in cases when data scale increases at intervals and hardware computational resources are limited.

Our main goal is to conduct an experimental evaluation of models namely, BERTopic, (Bidirectional Encoder Representations from Transformers)
\cite{grootendorst2022}, 
LDASequence (Latent Dirichlet Allocation Sequence) \cite{blei2006, dieng2019}, 
Top2Vec \cite{angelov2020}, 
and HDP (Hierarchical Dirichlet Process) \cite{Paisley2014} 
in dynamic topic modeling, using COVID-19 tweets from January, 2020 to December, 2021. We will compare each of these using evaluation metrics namely topic density, quality, evolution, stability, and execution time, and 
we implement the computation of evaluation metrics namely evolution and stability to track changes in topics when dealing with an event-driven process. The findings from this investigation will not only help to demonstrate the strengths and limitations of each model, but it will also offer research-based guidance on which algorithm the research communities should select in certain conditions where topics that evolve with time are studied. 

\textbf{Contributions.}
We present a methodology to evaluate dynamic topic modeling algorithms including selected evaluation metrics. Also, we prepare the evaluation dataset from COVID-19 tweets from January, 2020 to December, 2021. In addition, we conduct an extensive analysis of four selected algorithms, compare them, and present our findings.
\section{Topic Modeling}
\label{sec:topic-modeling}
Topic modeling is a concept that unveils the themes found in a collection of documents \cite{blei2006}. 
It 
is employed in numerous fields to represent vast volumes of digital text from computer and online technologies in a more manageable format.
\begin{figure}
    \centering
    \includegraphics[width=1.0\linewidth]{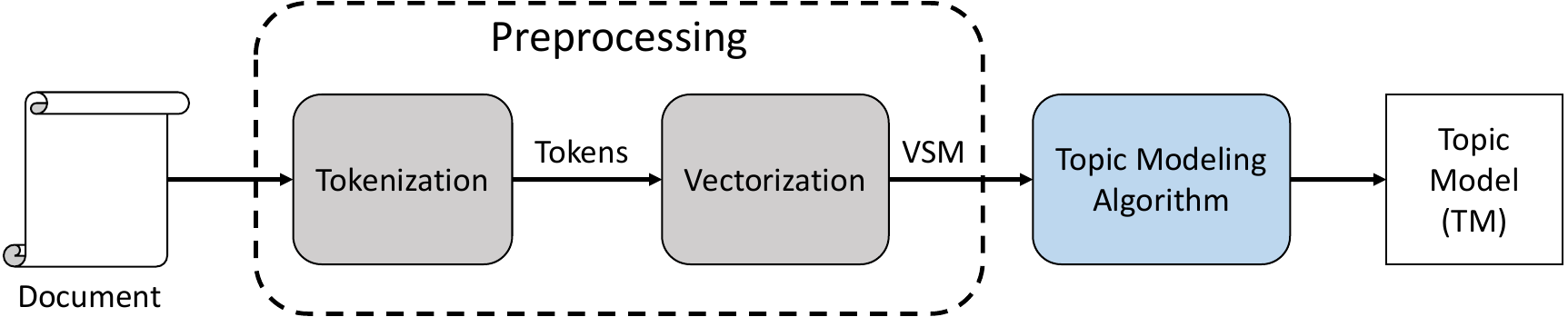}
    \caption{Stages of topic modeling.}
    \label{fig:topic-modeling}
\end{figure}
Applications of topic modeling can be found in fields such as bio-informatics (e.g., gene function annotation), economics (e.g., policy review), social sciences (e.g., history text mining), and Natural Language Processing (NLP) (e.g., text analytics) \cite{azzaakiyyah2023, liu2016, vijayan2021, valdez2018, de2021}. 
In the following, we explain how topic models (TMs) are computed. See the stages for building them in Figure \ref{fig:topic-modeling}.

\textbf{Tokenization}.
Tokenization is considered a very important step in NLP and text analysis. It is the process of dividing texts into smaller sizes often called tokens (e.g., words, phrases, sub-sentences,) \cite{domingo2018, toraman2023}. 

\textbf{Vector Space Models (VSM)}.
Topic modeling algorithms receive VSMs as input. Thus, the idea of topic modeling is based on the vector space model (VSM) which is used for information retrieval. 
VSM is an algebraic model that collects semantic data about word usage that is number representation of sequence of words \cite{Pavithra2024a}. 
Example of Vector Space Models are as follows-:
\begin{enumerate}
    \item Bag-of-words: BoW prioritizes word occurrence over word sequence. And it is the most widely used method for converting words into a vector space \cite{egger2022b}.
    \item N-gram/Bi-gram: captures some degree of context and word order by grouping words that are close to each other \cite{lahkar2022}.  
    Word sequence's integration in topic modeling should be taken into account using n-gram and bi-gram statistics \cite{wallach2006}. 
    \item \textit{Term frequency-inverse document frequency (TF-IDF)}: TF-IDF  
    is responsible for the term-document matrix where the importance of terms discovered in documents are determined by giving more priority to terms that did not occur frequently and less priority to terms that did occur more frequently \cite{kim2019, egger2022b}. 
    \item Word/Document Embeddings: as an alternative, \textit{word embedding} like GloVe, FasText \cite{badri2022},
   and Word2Vec 
\cite{jatnika2019} capture the semantic meaning between words. While \textit{document embedding} like Bidirectional Encoder Representations from Transformers 
\cite{grootendorst2022}, Universal Sentence Encoder 
\cite{cer2018}, and Doc2Vec 
\cite{lau2016} focus on capturing the meaning and perspective of entire documents. 
\end{enumerate}


Techniques (i.e. algorithms in this paper) employ sophisticated computational techniques to understand human languages as texts and cluster-related terms as topics and, thus, to fit models (e.g., TMs) on data (e.g., VSM) and use these models for estimations. 

\textbf{Algorithms.} 
Here, we present topic modeling algorithms.

\textbf{Latent Dirichlet Allocation (LDA):}
For a given corpus, LDA requires a predetermined number of topics. 
Being a generative model, LDA does not account for word semantics. It takes into account the probability distribution of variables and Bag-of-words \cite{kherwa2019, blei2006}. 
Since LDA supports multiple memberships, many topics can be assigned to a document. 
To obtain an ideal generated list of topics, its hyper-parameters must be properly adjusted, which is one of LDA's drawbacks \cite{egger2022a}. 
Furthermore, the deterministic character of the topics and their validity are not guaranteed due to the stochastic nature of the model \cite{egger2022a}. 

\textbf{Hierarchical Dirichlet Process (HDP):}
HDP automatically finds the number of topics.
HDP presents the topics it extracts in a top-down order (a tree-like structure) \cite{Paisley2014}. The inferential target is that general topics are at the top 
and abstract topics at the bottom. Furthermore, as the number of topics is unknown in advance or is anticipated to change over time, HDP is far more convenient, due to the non-parametric structure. 

\textbf{Top2Vec:}
Top2Vec uses document embedding \cite{angelov2020}
to semantically relate words to be in the same space during vectorization \cite{egger2022b}. 
Words like \textit{"computer"} and \textit{"programming"} for example, should be more similar than words like \textit{"programming"} and \textit{"baby"}. 
Top2Vec uses universal manifold approximation and projection (UMAP) for dimension reduction \cite{mcinnes2018}. 
HDBSCAN \cite{ibraimoh2024}  
then performs clustering based on the detected dense regions in the documents' VSM. Top-words that are most closely related to documents and are thought to be extremely important in characterizing the topic of a given document are selected. Consequently, Top2Vec automatically determines the number of topics \cite{angelov2020}.  
Predefined models like \textit{'universal-sentence-encoder-multilingual'} \cite{yang2019} 
are used by Top2Vec as its embedding model to facilitate multilingual analysis. 
In addition, Top2Vec assigns a document to a single topic 
\cite{egger2022a}. 

\textbf{BERTopic:}
BERTopic employs a sentence-transformer model that covers around 50 languages and leverages BERT (Bidirectional Encoder Representations from Transformers) as a document embedding extraction technique \cite{grootendorst2022, Pavithra2024a} 
Also, it employs embedding models and UMAP. BERTopic identifies the number of topics on its own by using clustering algorithms like KMeans and HBDSCAN \cite{ibraimoh2024} 
to cluster documents into expected topics \cite{egger2022a}. 
The main difference between Top2Vec and BERTopic is that the latter uses the c-tf-idf class \cite{grootendorst2022} 
that computes tf-idf \cite{kim2019} 
for each class to weigh each token and build up a topic. 
A crucial aspect to consider is that BERTopic can produce outliers, which are consistently indicated as -1 and are not anticipated to be subjected to more research \cite{egger2022a}. 
It assigns each document to a single topic and generates more topics than Top2Vec, making it more difficult for topic readability \cite{egger2022a}. 

\subsection{Dynamic Topic Modeling}
\label{section:Dynamic}
An extension of topic modeling in which topics are tracked over time is called dynamic topic modeling. It takes into account the temporal dependencies that are associated with data 
and so captures the evolution of the topics across time \cite{Pavithra2024a, blei2006}. 

This is particularly useful in cases when the dynamics of the events that are understudied entail changes. For example, records like emails, papers, research journals, and even social media trends \cite{blei2006} 
such as COVID-19 tweets, tend to change in time like the pandemic break out in other countries (e.g., Germany in March 2020 and vaccination campaigns in November 2020). 

In the following, we briefly explain dynamic topic modeling approaches.

\textbf{LDASequence:} LDASequence is designed to build DTMs. Time slices are incorporated into the modeling process of LDA \cite{Pavithra2024a, egger2022a, kim2019}. LDASequence works as follows: 
\begin{enumerate}
    \item Segmentation: LDASequence divides documents into time slices (e.g., days, months, years, quarterly) and each time slice comprises of documents within the specified period. LDA model is trained for each time slice to get topics, these models further allows topics to change over time by sharing information among time slices.  
    \item Information Sharing: information sharing is done by linking topic distributions between consecutive time slices. This is achieved by using the output of LDA model from the previous time slice to influence the initialization of the LDA model for the next time slice. Thus, a chain of dependency that captures the topic evolution is actualized. To ensure that temporal dependencies are regarded, methods such as variational inference, Gibbs sampling are used for topic distributions.
    \item Regularize Topics: regularization is then applied between time slices to moderate the changes of words that topic comprises. 
\end{enumerate}

\textbf{BERTopic:}
 BERTopic 
 calculates the topic's top words without repeatedly generating the whole model. The steps are as follows \cite{grootendorst2022}: 
 \begin{enumerate}
     \item Global Representations: 
     BERTopic finds topics for the whole texts, ignoring the temporal granularity (e.g., months, years). 
     \item Topics/Timestamps: 
     global representations of the texts are then assigned to respective temporal granularity.
     \item Compute c-TF-IDF: 
     for every temporal granularity with its associated topics, c-TF-IDF representation is computed and topic representations for every temporal granularity is created without repeatedly generating the whole model from scratch.
 \end{enumerate}

\textbf{HDP and Top2Vec:}
In a dynamic 
environment, 
TMs can be also used.
The steps are as follows
\begin{enumerate}
    \item Segmentation: Divide the documents into time slices.
    \item HDP/Top2Vec Model: for each time segment, apply HDP or Top2Vec model independently to find topics within that time segment.
    \item Match Topics: from each time slice, get and align topics across consecutive time slices to track their topic evolution. This is achieved by measuring similarities (e.g., cosine similarity metrics, jaccard similarity score) between topics across time slices. 
\end{enumerate}


\subsection{Evaluation Measures for TMs and DTMs}
The most popular evaluation metrics for TMs and DTMs are shown in this section. 
\subsubsection{Topic Coherence}
\label{section:coherence}
Topic coherence captures the semantic relationship between the topic words \cite{james2023}. 

First, the term probability is calculated:
\begin{align}
    P(w_a) = \frac{count(w_a)}{{|\beta|}}
\end{align}
where $count(w_a)$ represents the total number of documents where a word $w_a$ is found, $a$ refers to the index position of the word in a topic, and $|\beta|$ represents the total number of documents. 

The co-occurrence probability is computed by observing two words $w_a$ and $w_b$ that belong to the same document.
\begin{align}
    P(w_a,w_b) = \frac{count(w_a,w_b)}{{|\beta|}}
\end{align}

A point-wise mutual information PMI is calculated as follows: 
\begin{align}
    PMI(w_a,w_b) = \log \frac{P(w_a,w_b)}{P(w_a)P(w_b)}\label{equ:PMI}
\end{align}
PMI is normalized as follows: 
\begin{align}
     NPMI(w_a,w_b) = \frac{PMI(w_a,w_b)}{-\log P(w_a,w_b)}\label{equ:stability} 
\end{align} 
   
\textbf{{\textit{Variations of Topic Coherence}}}
\begin{itemize}
    \item $C_{UMASS}$ coherence score is non-normalized coherence that is based on co-occurrence counts of pairs of words. But in this case, the pairs of words come from a given sliding window that can be found within the documents
    \item $C_{NPMI}$ 
    relies on normalized point-wise mutual information among pairs of words.
    \item $C_v$ coherence score is calculated by combing indirect cosine matrix with NPMI. 
\end{itemize} 

Different variants of topic coherence were used to assess models in an event-driven setting \cite{montero2023} 
while \cite{james2023} 
presented \textit{Temporal Topic Coherence (TTC)} for dynamic topic models, which takes into account the topic coherence of word pairings that belong to a single topic across time slices. 
\subsubsection{Topic Diversity}
Topic diversity is seen as how frequently words occurred across all topics obtained from a model \cite{burkhardt2019}. 
When the topic diversity is approaching zero, it is said that topics contain words that are not found in any other topic while when approaching one it shows that words found in a topic are often seen in other topics. Dieng et. al \cite{dieng2019} 
computed topic diversity involving the number of unique words found in all topics divided by the total number of words found in all topics. It simply means that topic diversity approaching one shows that the topics are distinct enough.
In dynamic topic modeling context, \textit{Temporal Topic Smoothness (TTS)} for calculating the diversity of topics across time slices is captured \cite{james2023}. 
\subsubsection{Topic Quality}
 Topic quality is a combination of topic coherence and diversity. \cite{james2023} proposes a topic temporal quality for the DTMs. 
 High TTS and low TTC merely indicate that there has been a shift in the vocabulary employed throughout time. Low TTS and high TTC 
 demonstrate an increase in the topics' semantic relationship.

\section{Related Work}
The majority of researchers concentrated on comparing the qualitative and quantitative performances of different models. To examine Twitter posts, \cite{egger2022a} 
compared LDA, NMF, Top2Vec, and BERTopic. They used two distinct datasets,
\textit{"flight"} and \textit{"travel bubble"} to analyze these models. The models were compared using a qualitative analytical approach, in which the researchers assessed the topics they had selected using descriptive observations. 
Optimized LDA topic, base LDA models, and BERTopic model were examined on texts from \textit{X} platform \cite{montero2023}. The tweets included postings about specific areas in Metro Manila, Philippines. In order to do a quantitative analysis, coherence scores in Section \ref{section:coherence} 
for LDA and BERTopic were computed, including perplexity scores for LDA model. Optimized LDA outperformed the other models. Pavithra and Savitha \cite{Pavithra2024a} 
compared LDA, HDP, NMF, BERTopic, and DTM for evolving textual data on research papers. The main emphasis is on the difficulties involved with dynamic topic modeling. They demonstrated that when it comes to finding topics in research trends, the DTM performs better. 
Furthermore, f1-score, precision, and recall are used as evaluation criteria. Pavithra and Savitha \cite{Pavithra2024b} proposed a hybrid topic model that is extended to integrate temporal dynamics of topics for real-time and evolving textual data called CT-DTM models. It was compared to LDA, DTM, GIBBSLDA++, and HDP. Evaluation criteria like f1-score, accuracy, recall, precision, coherence, and perplexity were employed to assess the models and CT-DTM outperformed other models.

The evaluations done by \cite{egger2022a} 
is based on qualitative analysis subjected to only human interpretations. When the models were evaluated quantitatively \cite{Pavithra2024a,Pavithra2024b}, a detailed description of how the topics evolved was missing. In other words, previous researchers compared either between static topic modeling algorithms or a static topic modeling algorithm against a dynamic topic modeling algorithm. On the contrary, our work investigates how the topics evolve by focusing on an empirical evaluation of and comparison between dynamic topic modeling algorithms, thereby addressing the research gap outlined earlier. 
In an event-driven case as seen in \cite{Pavithra2024b, montero2023}, the relationships between past and present topics after an event occurred are not captured. Our contribution is targeted at extensively comparing DTMs using evaluation metrics: topic density, quality (coherence and diversity), topic evolution, topic stability, and execution time. In this regard, we 
implement how quantitatively the relationships between topics can be computed in an event-driven environment and additionally study the DTMs' behavior of scaling up the data in a limited hardware resource platform. 
\section{Evaluation Methodology}
\begin{figure}
    \centering
    \includegraphics[width=1.0\linewidth]{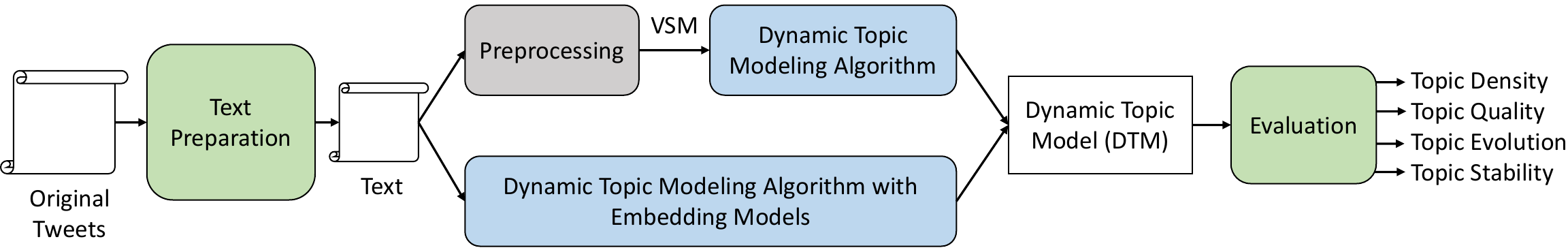}
    \caption{Evaluation methodology.}
    \label{fig:methodology}
\end{figure}

Our evaluation methodology lies in preparing the evaluation dataset (i.e. text) and the evaluation process, both shown as green rounded rectangles in Figure \ref{fig:methodology}.
Text preparation (Section \ref{subsec:text-preparation}) includes data collection, data cleaning, pre-processing, and text sample selection. 
In Section \ref{subsec:implementation-of-evaluated-algorithms}, we explain in detail the changes we make in each algorithm we selected for evaluation purposes.
In Section \ref{evaluationM}, we present the evaluation metrics we used for evaluating and comparing the selected algorithms. 
\subsection{Text Preparation}
\label{subsec:text-preparation}
We collect the tweets 
from Germany created from January, 2020 to December, 2021 related to COVID-19. The data collection was done using the Twitter API based on a manually collected list of government accounts, https://zenodo.org/records/14046854. The tweets have been tagged for COVID-19 related content by Twitter (now X) itself and the tags could be retrieved via the API. We verified the quality of the data in extensive evaluations that is not part of this paper. 
The COVID-19 tweets datasets have around 725,414 tuples in total.
Each tuple contains the text of a tweet or a comment (i.e. the text field). 
We filtered texts written in German, resulting in 691,875 tuples. This will also support the model's performances in another language apart from English. 
These algorithms require huge computational resources to build TMs. To be able to include these algorithms in our evaluation, we reduce the size of the dataset to 107,011 (called evaluation dataset (twitter)) by random sampling without replacement, considering having a representative sample of each month. As we will show later, even with 107,011 tuples, difficulties in execution of some models is observed.

To measure the scalability of the models as the data increases in a limited hardware resource environment, 
, we split each tuple that has a huge text into multiple tuples using NLTK tokenizer \cite{nltk}, each tuple stores a sentence. With this, the length of vector space per text is reduced.


Our target is to extract the location (i.e. the latitude and the longitude as a filter criteria) from the text field in our dataset.
In order to accomplish this, we use the pre-trained model \textit{"de-core-news-sm"} in the Spacy package \cite{SpacyDe} for named entity recognition \cite{keraghel2024}, that assigns words to various entities. An example of an entity is "Person", "\text{}Organization", "Time", "\text{}Location" \text{} etc.
In this instance, we are able to obtain entities that are classed as locations (called "LOCs"). Despite the pre-trained model's f1-score, precision, and recall of 0.82, 0.83, and 0.81, respectively, in detecting entities, a few false classifications of terms as "LOC" \text{} were observed. We double-check the locations extracted by including human-in-the-loop and, then looking up the longitude and latitude of these locations using the GeoPy framework \cite{GeoPy}. We come to the conclusion that an entity has been incorrectly classed as a "LOC" \text{} if the location's longitude and latitude retrieved are zeros. Thus, the tuple is discarded. 
As a result of this stage, each text field in a tuple is associated with an extracted location. Thus, grouped texts by location will be used for scalability testing.

The text fields of the resulting dataset include stop words, 
emojis, special characters, etc. Therefore, data cleaning is an essential step in preparing texts. Because the most important terms are used when learning the topics.
We used RE \cite{Re} and NLTK \cite{nltk} frameworks for data cleaning. 

\begin{figure}[t]
    \centering
    \includegraphics[keepaspectratio,height=12cm, width=8cm]{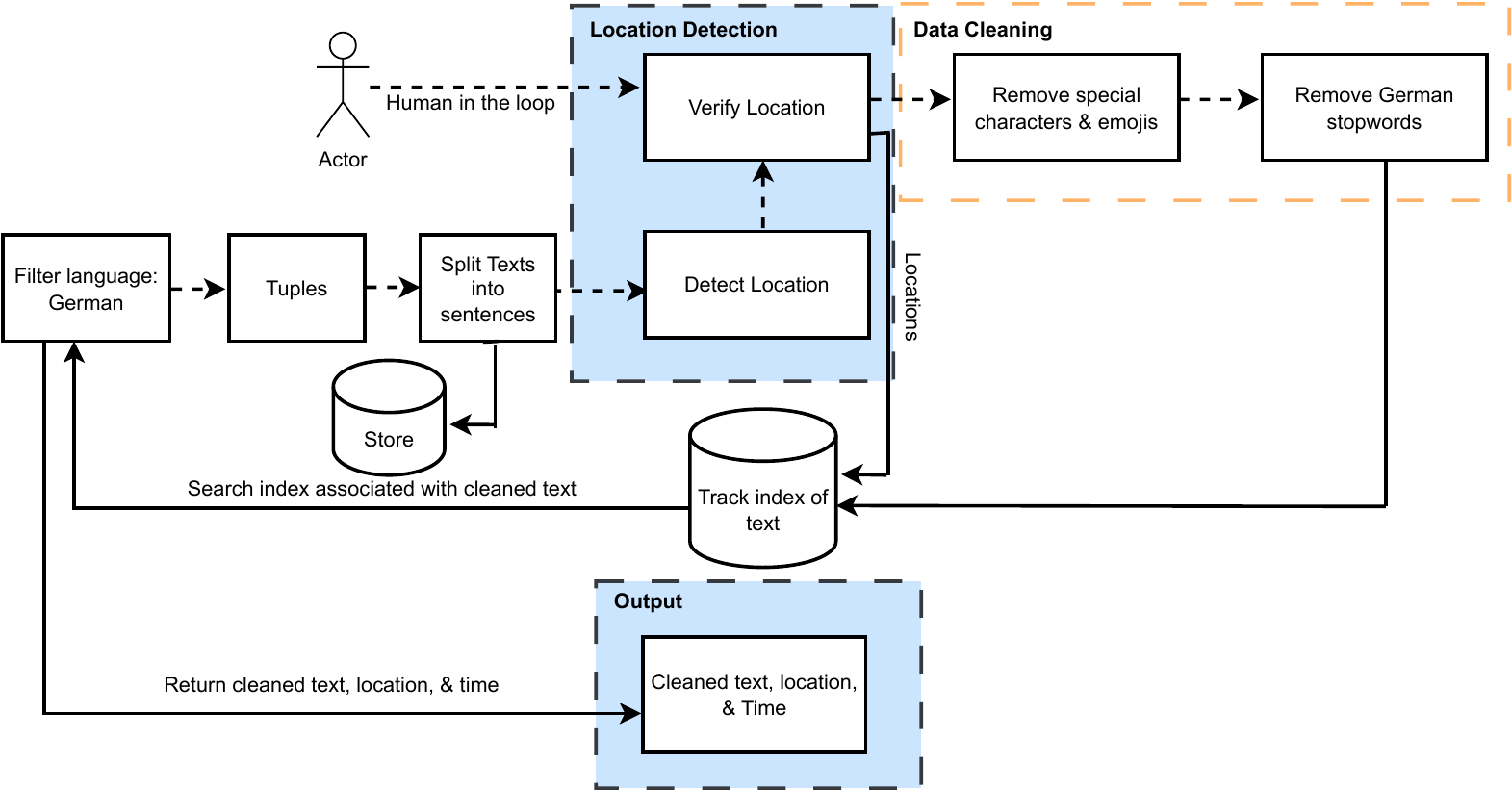}
    \caption{Stages of text preparation.}
           \label{fig:Data_Cleaning_Process}
\end{figure}

For scalability tests, we filtered the four top locations with a large number of texts among all locations namely Germany, France, Italy, and China. This results in the \textit{evaluation sub-dataset}.
All the text preparation stages mentioned above are depicted in Figure \ref{fig:Data_Cleaning_Process}.
We further evaluate scalability of the models using a different dataset called \textit{UN-Debate datasets}, collected from \cite{kaggle} with timestamps from year 2003 to 2015. Also, an overall evaluation is provided for UN-Debate datasets with timestamps from year 1970 to 2015.


\subsection{Configuration of Evaluated Algorithms}
\label{subsec:implementation-of-evaluated-algorithms}
For our evaluation, we select two algorithms from the probabilistic model class (namely LDASequence and HDP) and two algorithms from the embedding models one (namely BERTopic, and Top2Vec). We chose these algorithms captured in  related work to show that they can also be extensively evaluated with our implemented evaluation metrics. 
We develop an incremental procedure such that each algorithm trains its TM incrementally month by month starting from January 2020 using a \textit{month slice} from the dataset.  Month slices can also be replaced by any other time slices depending on the temporal granularity (e.g., year, quarterly, day).  
As shown in Figure \ref{fig:methodology}, the algorithms that use embedding models include tokenization and vectorization. Therefore, month or year slices with texts are directly the input for BERTopic and Top2Vec while preprocessing 
is required for LDASequence and HDP.


\subsubsection{LDASequence}
For LDASequence algorithm, the main parameter is the number of topics that we select experimentally. 
Using TF-IDF representations, we transformed the tokens of every dataset into vectors which is the input for LDASequence. The first step in using LDASequence is to sort and group the tuples based on months (or years), and count the tuples of each month slice. The LDASequence then uses this counted months and vector space (term document matrix) determine how the topics changed over time as described in Section \ref{section:Dynamic}. LDASequence operates under the premise that a text might be associated with many topics. We identify texts related to topics and capture the maximum probability score of topics assigned to each text. 
\subsubsection{HDP}
We also use TF-IDF 
for HDP algorithm.
For each tuple, HDP (explained further in Section \ref{section:Dynamic}) associates multiple topics to the text of the tuple. 
. 
As we do for LDASequence, we weigh topics considering those with the highest probability to be associated with the respective text. 
We further assign respective timestamps to the texts. 
\subsubsection{BERTopic}
We select \textit{"distiluse-base-multilingual-cased-v1"} sentence transformer \cite{cer2018, reimers2019} to be the one that BERTopic uses for vectorization. 
As mentioned in Section \ref{sec:topic-modeling}, UMAP and c-tf-idf models are used for dimension reduction and class-based term-document matrix representations respectively. 
After fitting BERTopic model, we send the list of month (or year) slices to topic-over-time component of BERTopic which assigns the topics to respective timestamps and specifically assigns timestamps to each related text. 
\subsubsection{Top2Vec}  
The word embedding transformer included in Top2Vec is \textit{universal-sentence-encoder-multilingual} \cite{yang2019}. 
Top2Vec requires a huge dataset during initialization. We therefore doubled the original texts of the first month (or year) before performing the remaining incremental procedures. It produces document-id, which are eventually assigned to texts along with the timestamps that correlate to these texts. Explained further in Section \ref{section:Dynamic}. 
\subsection{Selected Evaluation Metrics}
\label{evaluationM}
To demonstrate the goal of tracking how topics change when an event occurs, we present here how topic evolution and stability 
are generated in an event-driven environment are computed. We also profile how the topic density and topic quality of each DTM are computed alongside the time required to build them i.e. the execution time of the algorithm.
\subsubsection{Topic Density}
The percentage of texts that each of the identified topics represents is captured by topic density. 
We calculate 
topic density as the total number of topics divided by the total number of texts. This helps in recognizing the growth in the number of topics. 
With more topics, the topic density tends to increase and
conversely, with fewer topics, the topic density decreases.
The higher the density, the better for cases where the idea is to get all the hidden topics in a text while the lower the density is, the better for readability of the topics.
 \subsubsection{Topic Quality}
Topic quality includes topic diversity and coherence. As we update the models, topic diversity captures the distinctions or originality between topics extracted at each iteration. 
Whereas topic coherence documents the interpret-ability of the topics extracted. 

Topic diversity is computed by dividing the total number of distinct words across all topics by the total number of word unions across all topics. 
The type of coherence that we used is $C_v$ described in Section \ref{sec:topic-modeling} and is computed using Gensim framework \cite{gensim}. 
\subsubsection{Topic Evolution}
\begin{figure}[h]
    \centering
\includegraphics[width=0.8\textwidth]{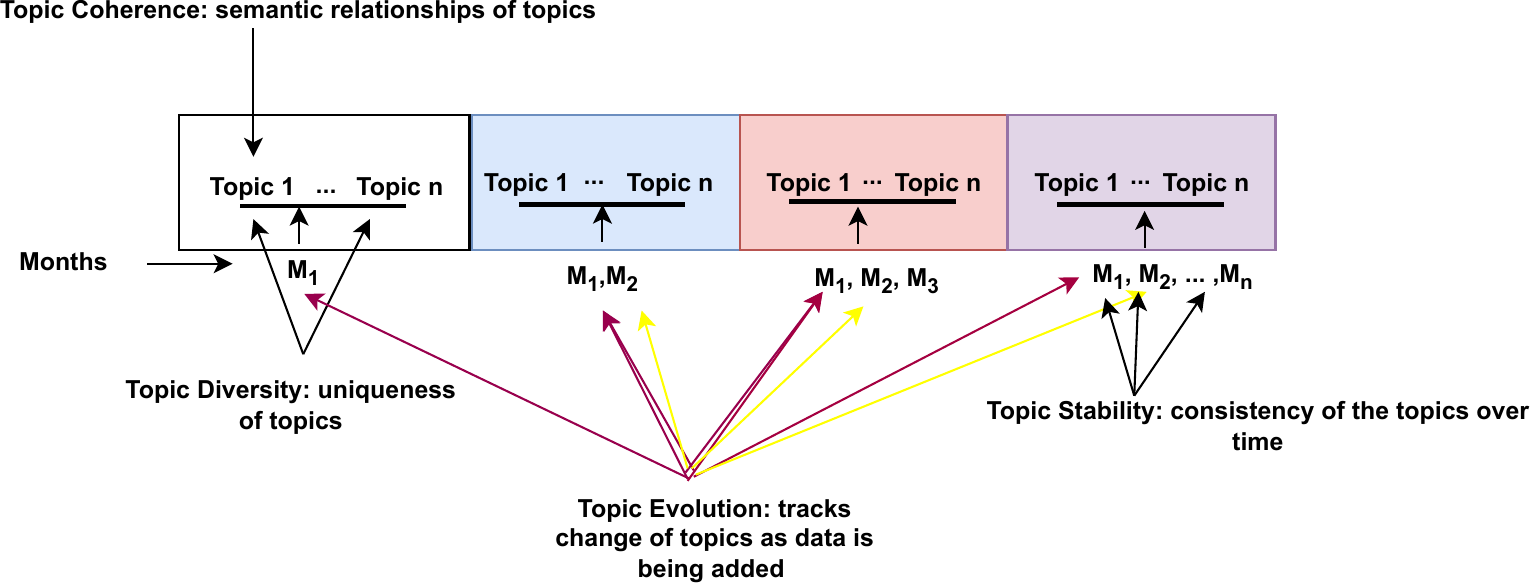}
    \caption{Application of evaluation measures in the iterative process of updating the models.}
    \label{fig:tikzpgf}
\end{figure}

The main difference between topic evolution and topic diversity is that topic evolution compares the uniqueness of topics found for a given month to all topics assigned to the same month, as shown in Figure \ref{fig:tikzpgf}, after an increment of the model with texts from the following months. 
In this instance, the interpretation of topic evolution is the reverse of topic diversity (i.e., $1 - topic diversity$). 
In Figure \ref{fig:tikzpgf}, the topic evolution of the topics of a month ($M_1$) before adding and after adding time slice of the following month ($M_2$) is between topics of $M_1$ in white and $M_1$ in blue. 
For example, value 0.2 means that the topics discovered before adding $M_2$ and after the addition are similar. The lower values mean the similarity of topics is higher.
This aids in capturing patterns in the relationships of topics allocated to a specific month during model updates, as well as the number of topics discovered. We calculate 
topic evolution as follows:
\begin{align}
\centering
T_{Evol} = \frac{|\bigcup^{T_{{M_i{_p}},{M_i{_c}}}}_{n=1,i=1} W_n|}{T \cdot N}\label{equ:evolution}
\end{align}
\begin{itemize}
    \item ${T_{{M_i{_p}},{M_i{_c}}}}$ represents all topics before adding a month slice ($M_i$) and after the addition.
    \item $W_n$ represents 
    top-n-words before adding a month slice ($M_i$) and after the addition.    
    \item $|\bigcup^{T_{{M_i{_p}},{M_i{_c}}}}_{n=1,i=1} W_n|$ represents the number of unique words $W_n$ found before adding the month slice of $M_i$ and after the addition.    
    \item $T \cdot N$ represents the number of unions of $T$ top-n-words before adding the month slice of $M_i$ and after the addition.
    \item If $T_{Evol}$ $\rightarrow$ $1$, we say that the topics $T_{M_i{_p}}$ and the topics $T_{M_i{_c}}$ are not closely similar. 
    \item If $T_{Evol}$ $\rightarrow$ $0$, we say that the topics $T_{M_i{_p}}$ and the topics $T_{M_i{_c}}$ are closely similar. 
\end{itemize}

Equation \ref{equ:evolution} captures how the topic changed during the incremental process. 
\subsubsection{Topic Stability}
\begin{figure}[h]
    \centering
    \includegraphics[width=0.3\textwidth]{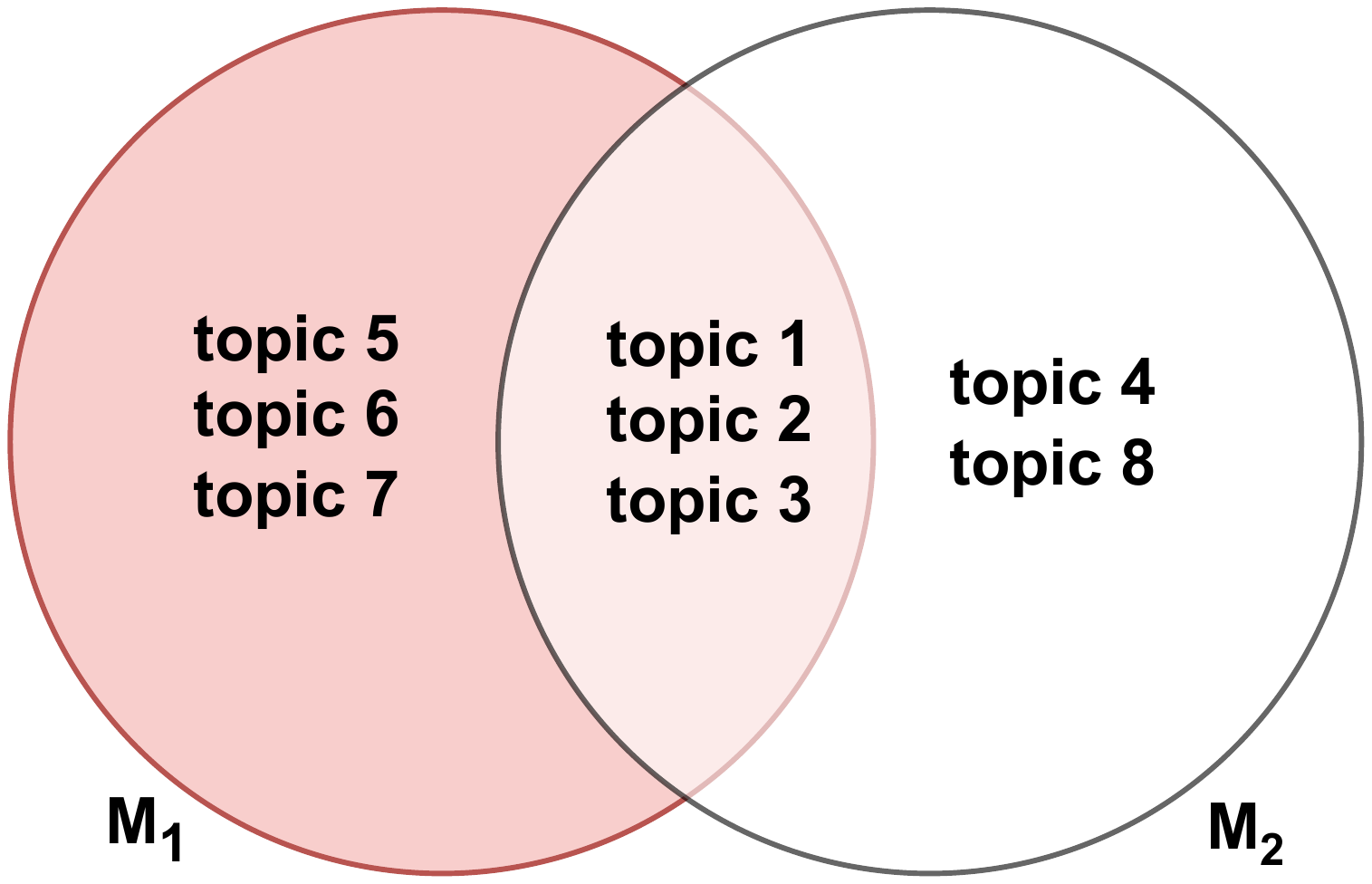}
    \caption{Topic stability.}
\label{fig:VenDiagram}
\end{figure}

Topic stability checks topic consistency across months at the end of the incremental procedure. To determine the relationship across the topics at different months as shown in Figure \ref{fig:VenDiagram}, we employed the Jaccard Similarity Matrix \cite{Fletcher2018} which divides the number of intersecting topics from two months by the number of union of topics from the same two months. This supports analyzing the relationships between topics over time from all the models considered. We calculate the topic stability of a month ($M_i$ for example) as follows:
 \begin{align}
T_{Stab} = \frac{|T_{M_i}\cap T_{M_j|}}{|T_{M_i}\cup T_{M_j|}}\label{equ:stability}
\end{align}
\begin{itemize}
    \item $T_{M_i}$ represents set of topics at month $M_i$, 
    \item $T_{M_j}$ represents set of topics at another month $M_j$.
    \item $|T_{M_i}\cap T_{M_j}|$ represents the number of intersecting topics between months $M_i$ and $M_j$.
    \item $|T_{M_i}\cup T_{M_j}|$ represents the number of union topics in months $M_i$ and $M_j$. 
    \item If $T_{Stab}$ $\rightarrow$ $1$, we say that $T_{M_i}$ and $T_{M_j}$ have higher similarity .
    \item If $T_{Stab}$ $\rightarrow$ $0.5$, we say that $T_{M_i}$ and $T_{M_j}$ have moderate similarity.
    \item If $T_{Stab}$ $\rightarrow$ $0$, we say that $T_{M_i}$ and $T_{M_j}$ are not closely similar.    
\end{itemize}

\section{Results}
In this section, we run the selected four dynamic topic modeling algorithms (Section \ref{subsec:implementation-of-evaluated-algorithms}) to extract topics for the dataset we prepared (Section \ref{subsec:text-preparation}) and compare their results using the evaluation metrics of (Section \ref{evaluationM}). 

We represent time slices in numbers such that the index of the first time slice $TS_1$ (i.e. January 2020 or 1970) is 1 and that of the last time slice $TS_{24}$ (i.e. December 2021) is 24 for twitter (covid-19) datasets. $TS_1$ (i.e. 1970) is 1 and $TS_{46}$ (i.e. 2015) is 46 for un-debate datasets.
The reason for using one month as a time slice for twitter datasets is because the number of tuples belonging to one month's time slot at every iteration is quite high and will reflect the event-driven environment more than any other intervals in the temporal granularity. Models need more tuples, for example Top2Vec at first iteration as explained in Section \ref{subsec:implementation-of-evaluated-algorithms}.
\subsection{Experimental Setup}
As mentioned earlier, the evaluation sub-dataset (twitter covid-19) is represented by four top locations, namely Germany, France, Italy, and China. See Table \ref{table:LocationDataDescription} and notice that location Italy have a different interval. This means Italy as a location was not mentioned in texts associated to January, 2020. We further picked recent years (totaling 13 years) from un-debate datasets. 
The hyper-parameters of the algorithms are tuned to capture the behavior of the models, as shown in Table \ref{table:ParametersSettings}.

\begin{table}[h]
  \centering
    \caption{
    Metadata of the evaluation dataset.}
\resizebox{0.8\textwidth}{!}{
\begin{tabular}{llll} 
 \toprule
 Datasets & Number of tuples & Interval & Tuple fields \\ [0.5ex] 
 \toprule
 \textbf{Twitter (covid-19)} & & &  \\
 Germany & 21,143 & 2020.01-2021.12 (24 months)& Year, Location, Text, Tweet-Id \\
 France & 1,337 & - & - \\
 Italy & 2,354 & 2020.02-2021.12 (23 months)& - \\
 China & 3,037 & 2020.01-2021.12 (24 months) & - \\
Evaluation dataset & 107,011 & - & - \\
\textbf{UN-Debate} & & & \\
UN-Debate dataset  & 2493 & 2003-2015 (13 years) & Year, Text, Country \\
Evaluation dataset  & 7507 & 1970-2015 (46 years) & Year, Text, Country \\
 \bottomrule
\end{tabular}
    \label{table:LocationDataDescription}}
\end{table}
\begin{table}[h]
  \centering
    \caption{Model parameters settings.}
\resizebox{0.8\textwidth}{!}{
\begin{tabular}{lllll} 
 \toprule
 Models & Hyper-Parameter & Evaluation sub-datasets & Evaluation datasets (Twitter) & Evaluation datasets (Un-debate) \\ [0.5ex] 
 \midrule
 BERTopic & min-topic-size & 2 & 100 & 20\\ 
 Top2Vec & min-count & 2 & 100 & 100 \\
 LDASequence & num-topics & 6 & N/A & N/A\\ 
 HDP & gamma, alpha, T & 0.1, 0.04, 5 & 1.0, 0.1,100 & 0.1, 0.04, 5\\
 \bottomrule
\end{tabular}
\label{table:ParametersSettings}}
\end{table}
\textbf{Hardware.}
We use an NVIDIA A100 Tensor Core GPU with 40GB of memory connected via PCIe 4.0 to a dual-socket AMD EPYC 7F52 processor with 16 cores, and CUDA 12.2.
\subsection{Topic Density}
\begin{figure}
    \centering
    \includegraphics[width=0.3\linewidth]{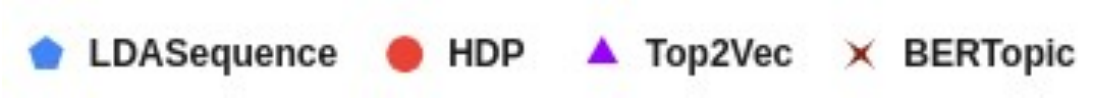}
    \includegraphics[width=0.9\linewidth]{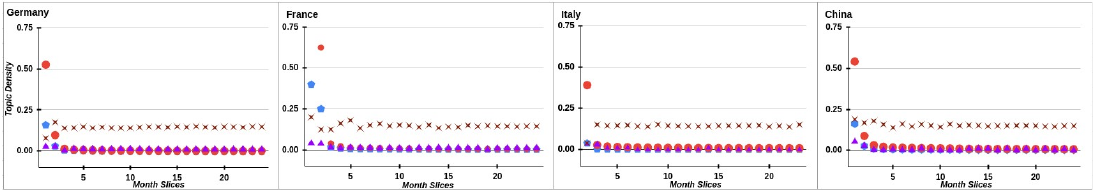}
    \includegraphics[width=0.7\linewidth]{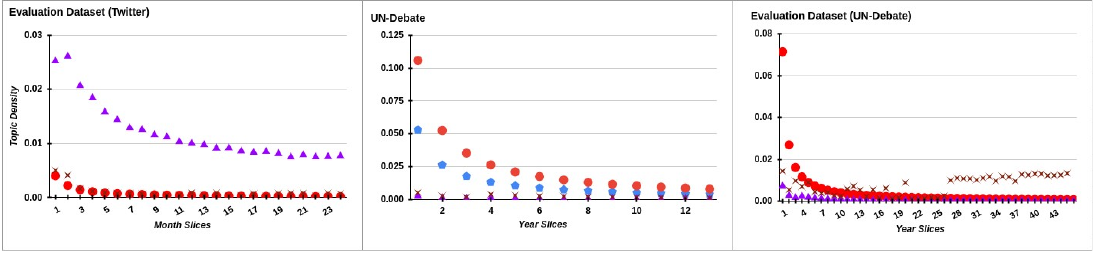}
    \caption{Evaluation of topic density of models.}
    \label{fig:TopicDensity}
\end{figure}

Figure \ref{fig:TopicDensity} displays the topic density of the models for each of the four selected countries, evaluation dataset (twitter), un-debate sub-dataset, and evaluation dataset (un-debate). 

As the amount of text increases, BERTopic generates a disproportionately large number of topics. For example, as of $TS_2$ and $TS_3$ in Germany, the total text is $1609$ and $2714$, and the number of topics produced by BERTopic is $224$ and $385$ respectively. At every increment, BERTopic consistently produced a higher number of topics than other models due to the low value setting of \textit{min-topic-size} hyper-parameter as shown in Table \ref{table:ParametersSettings}. HDP initially captured a larger percentage of texts than BERTopic, but it then started to drop because it consistently produced $5$ topics for each event. Since the LDASequence model's number of topics $6$ is fixed, the topic density decreases as the number of texts rises. Top2Vec has a poor topic density since the number of topics varies less in each iteration. For example, Top2Vec identified $6$ and $36$ topics for $TS_2$ and $TS_3$, respectively. Figure \ref{fig:TopicDensity} shows that these observations hold for all four countries.  HDP and BERTopic shows the same behavior for evaluation dataset (twitter) in Figure \ref{fig:TopicDensity} except for Top2Vec with high topic density. This is due to high optimal value of parameter \textit{min-count}. For un-debate, Top2Vec produces only 1 topic at every iteration (e.g., even during hyper-parameter tuning,  when min-count is low or high). That is why Top2Vec's topic density is very low compared to other models. While BERTopic produces topics more than Top2Vec, HDP produces highest number of topics ($20$), followed by LDASequence with 10 topics. For evaluation dataset (un-debate), Top2Vec found 1 topic at every iteration as well. 
However, as we will show later, LDASequence computed both evaluation datasets (twitter and un-debate) for days and was killed due to high hardware consumption. From these observations, we say that HDP captures the proportion of text well at the initial stage, BERTopic and Top2Vec shows consistent behavior in topic density, however, they produce topics that are not readable most times, as they are more in number while and LDASequence produce fewer varieties of topics that can be easily understood.

\subsection{Topic Quality}
\label{section:quality}
\begin{figure}
    \centering
    \includegraphics[width=1\linewidth]{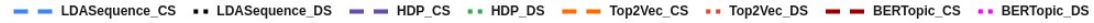}
    \includegraphics[width=0.98\linewidth]{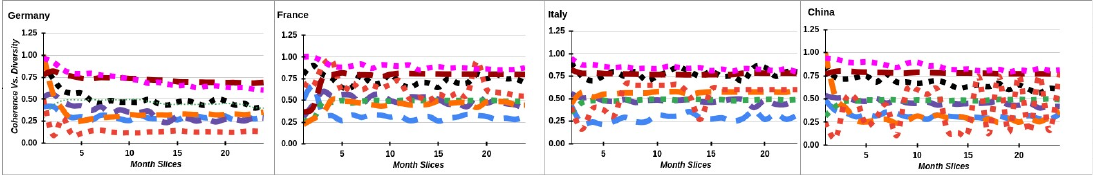}
    \includegraphics[width=0.77\linewidth]{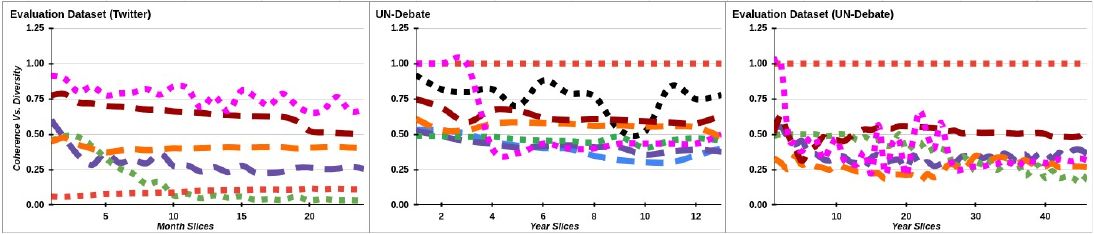}
    \caption{Evaluation of topic coherence and diversity of models performances on location, un-debate, and Evaluation datasets.}
    \label{fig:Coherence}
\end{figure}

For every update made on the models with the text of new time slices, we capture the topic quality (i.e., topic coherence and diversity ) as seen in Figure \ref{fig:Coherence}. As BERTopic generates a high number of topics, it is able to maintain high topic quality 
that converges. 
Therefore, for the average topic quality, BERTopic achieves a score of around $80\%$ except for the locations of Germany where it scores $71\%$.
On the other hand, the diversity of Top2Vec shows a tremendous rise and fall, especially at certain time intervals. 
This 
demonstrates that during some time slices, Top2Vec concentrates on fewer, more closely related topics in the cluster. 
 In Figure \ref{fig:Coherence}, the topic quality of Top2Vec for France, Italy, China, and Germany are  $54\%$, $54\%$, $36\%$ and $25\%$ respectively. From this, we can see that the topic quality of Top2Vec decreases with more number of tuples.

LDASequence has a low topic quality compared to BERTopic due to the high values of diversity score and low values of coherence across all countries. This behavior shows that words in topics are less related but they are more distinct from each other. This is also because the number of topics remains constant regardless of the number of tuples. 
However, its average topic quality is approximately $53\%$ for Italy and France except China and Germany with score of  $50\%$ and $40\%$ respectively. 
HDP's average topic quality $47\%$, except for Germany with a score of $40\%$, outperforms Top2Vec due to the constant extractions of the number of topics to be five by the algorithm automatically. HDP model tries to add more top words to topic number that is fixed. 
For un-debate dataset, BERTopic is still the best in topic quality with an average score of 59\%. However, the LDASequence and HDP average topic quality scores are 58\% and 44\% respectively. Top2Vec topic diversity as shown in Figures \ref{fig:Coherence} and \ref{fig:TopicDensity}, diversity and number of topics respectively (i.e., computing topic diversity for a single topic will always result to 100\% score) clearly shows that Top2Vec sees un-debate and evaluation datasets (un-debate) as a homogeneous datasets (i.e., very similar in contents). Hence Top2Vec model is not a good fit for a slightly homogeneous datasets.   

In consideration of the above evaluations, if the target is to achieve both topic interpret-ability and uniqueness, BERTopic should be preferred to LDASequence, HDP, and Top2Vec.
\subsection{Topic Evolution}
To be able to evaluate the topic evolution and stability extensively, 
we continue the evaluation on the whole evaluation dataset (twitter) rather than discussing each country separately as the general findings also apply to all countries. We omitted un-debate datasets for this evaluation due to the homogeneity nature of the dataset discussed in Section \ref{section:quality}.   

\begin{figure}
    \centering
    \includegraphics[width=0.8\linewidth]{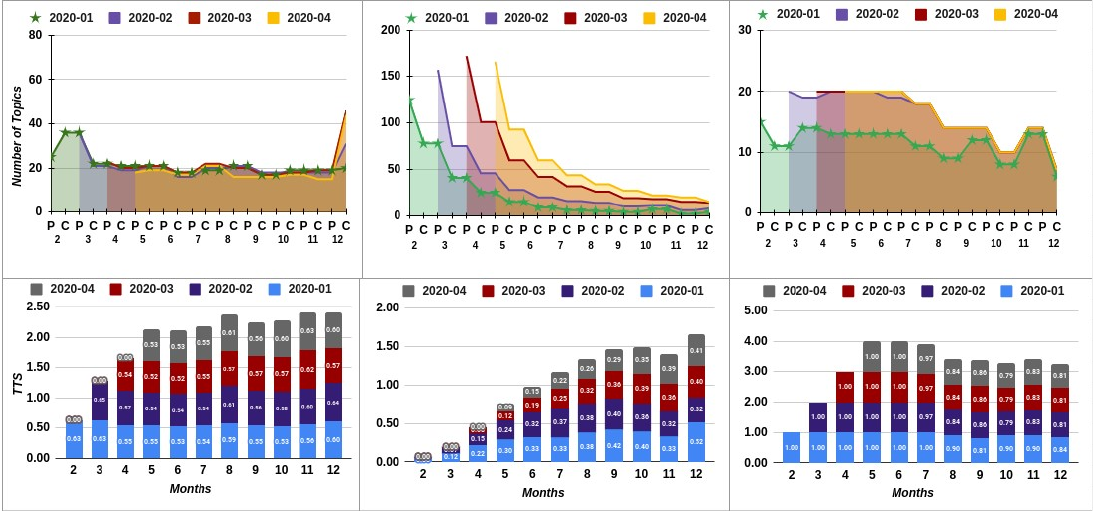}
    \caption{Evaluation of the number of topics and evolution of topics of models' performances. column 1: BERTopic, column 2: Top2Vec, column 3: HDP}
    \label{fig:TemporalTopicSmoothing}
\end{figure}

In Figure \ref{fig:TemporalTopicSmoothing}, we capture for the whole year of 2020 the topic evolution metrics 
that shows how the topics assigned to a particular month slice change. 
The changes noted here are the number of topics and the evolution, temporal topic smoothness (TTS) between the current and past topics of a month slice. 
Past topic length is referred to as \textit{P} while current topic length in the Figure \ref{fig:TemporalTopicSmoothing} is referred to as \textit{C}. 
The corresponding stacked histogram represents the TTS. 

During the incremental process, the TTS shows that Top2Vec and BERTopic tends to handle the evolution of topics assigned to a particular month slice better. 
For instance, the number of topics assigned to $2020-1$ in green in Figure \ref{fig:TemporalTopicSmoothing} column 1 before an update with texts from $2020-3$, represented by $P$, was $78$. 
After the update, the current topic length, represented by $C$, becomes $41$. Then the corresponding stacked bar plot shows that the past topics and current topics have a diversity score of 0.12, meaning they have $88\%$ similarity. However, the similarities decrease as updates are made with more month slices. Conversely, TTS of BERTopic shows that almost all the models update with texts from next month slices is around $52\%$ on average. 
On the other hand, HDP performs poorly in this phase despite optimizing the hyper-parameter $T$ that controls the maximum number of topics. At $T=150$, HDP still generates 20 topics. 
However, we found that topics are repeated. Thus, we filter the unique ones to avoid redundancy. That is why we have 20 topics as the maximum number of topics and 15, 9, and 8 in some time slices in Figure~\ref{fig:TemporalTopicSmoothing} column 3. The TTS between the past and the current topics of a month slice evolves drastically and tends to improve as the number of tuples grows. This is quite the opposite of the transition of TTS of Top2Vec and BERTopic.  

\subsection{Topic Stability}
\label{sectionstability}

\begin{figure}[h]
    \centering
    \hspace*{0cm}
    \includegraphics[width=0.42\linewidth]{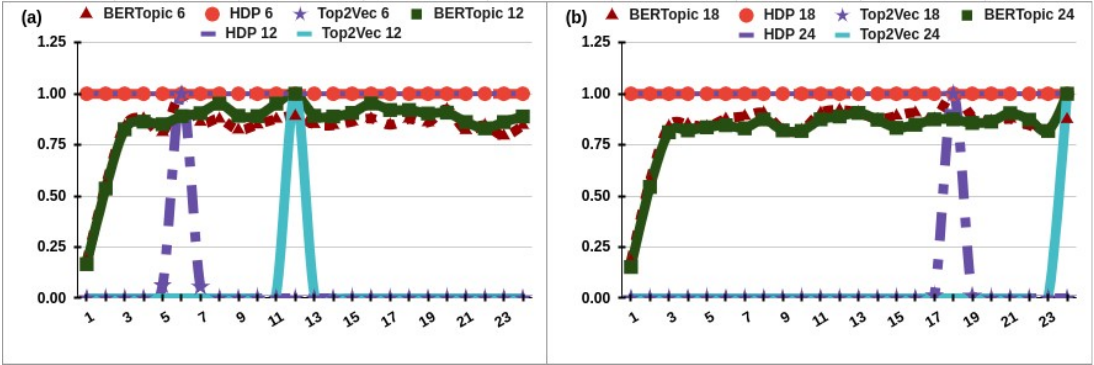}
    \includegraphics[width=0.45\linewidth]{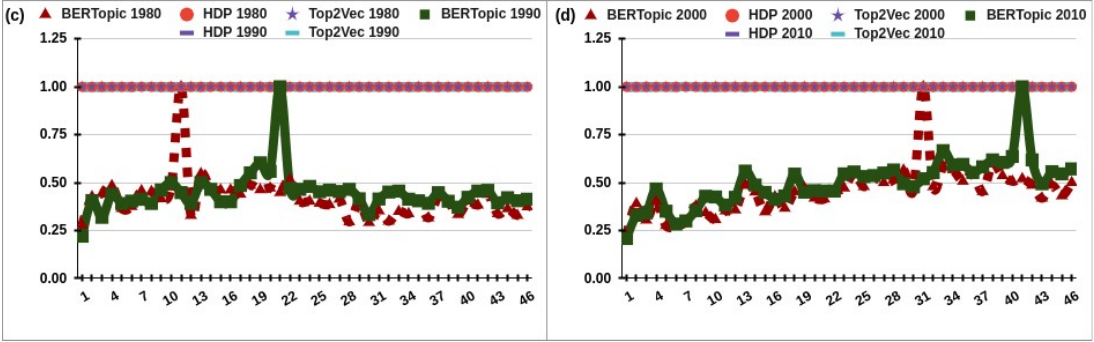}
    \caption{Topic stability. (a) shows 2020-06, 2020-12 and (b) shows 2021-06, 2021-12 topics comparison with other months for evaluation dataset (twitter). (c) shows 1980, 1990  and (d) shows 2000, 2010 topics comparison with other years for evaluation dataset (un-debate).}
    \label{fig:TopicStability}
\end{figure}

After updating all the models with the last month (i.e., $TS_{24}$) and year (i.e., $TS_{45}$) slices,   
we capture the relationships between the topics assigned to different months and years for the evaluation datasets: twitter and un-debate respectively.
Figure \ref{fig:TopicStability} shows Jaccard similarities between all models and the horizontal axis represents month and year slices.
Regarding the scalability limitation of LDASequence that we mentioned earlier, it is not executed for the evaluation datasets. 

We pick particular intervals of 6 months and 10 years for evaluation datasets: twitter and un-debate respectively. Figure \ref{fig:TopicStability} (a) and (b) show that we compare topics in $TS_6$, $TS_{12}$, $TS_{18}$, and $TS_{24}$ (i.e., where models with Jaccard similarity peak = 1.0 at the same time) with topics found in other months for evaluation dataset (twitter), while (c) and (d) show that we compare topics in $TS_{11}$, $TS_{21}$, $TS_{31}$, and $TS_{41}$ with other topics found in other years for evaluation dataset (un-debate).


For $TS_6$ represented in purple stars in Figure \ref{fig:TopicStability} (a), Top2Vec records a low Jaccard similarity. 
This implies that events that occurred before and after  $TS_6$ are not similar with events found in other months. The same behavior is observed for $TS_{12}$ and also in Figure \ref{fig:TopicStability} (b) when $TS_{18}$ and $TS_{24}$ are compared with other months. HDP at $TS_6$, $TS_{12}$, $TS_{18}$, and $TS_{24}$ shows high similarity score of 1.0 with topics from different months. Thus, the same event occurred all through the $24$ months. 
BERTopic topics in both $TS_6$ and $TS_{12}$ have a moderate relationship with $TS_1$ and $TS_2$.  But, subsequent months' relationships have a better Jaccard similarity score showing that different topics found for each month are related to topics found in $TS_6$ and $TS_{12}$. This is also applicable to $TS_{18}$, and $TS_{24}$ in Figure \ref{fig:TopicStability} (c).  This is realistic as
there are events that occurred at specific months during corona although some events can overlap into multiple months such as lockdown, vaccination, or even corona outbreak. 

For un-debate dataset in Figure \ref{fig:TopicStability} (c) and (d), BERTopic maintains the same trend. The major difference, is that the value of Jaccard similarity decreased. HDP maintains a high relationship, that is the Jaccard similarity scores are 
the same in all comparisons. 
However, Top2Vec have Jaccard similarity score of 1.0 between topics from other years. This behavior is different from our observation with evaluation dataset (twitter).

The above evaluation clearly shows that BERTopic presents better topic distributions among temporal granularity in an event-driven environment. 

\subsection{Performance}
\label{evaluation:performance}
\begin{figure}[h]
    \centering
\includegraphics[keepaspectratio,height=22cm, width=12cm]{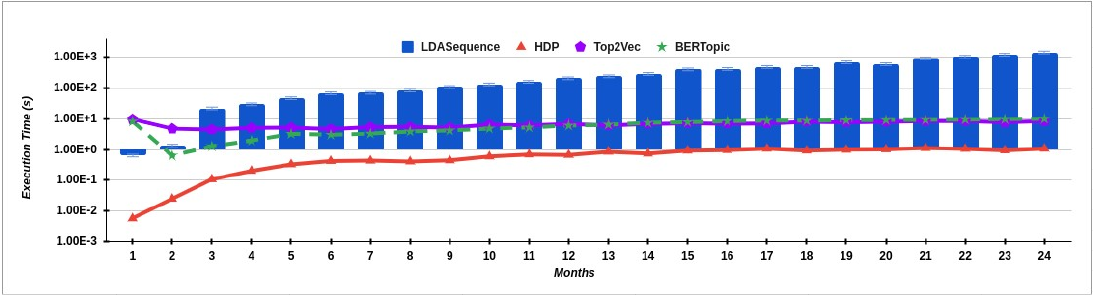}
\includegraphics[keepaspectratio,height=22cm, width=12cm]{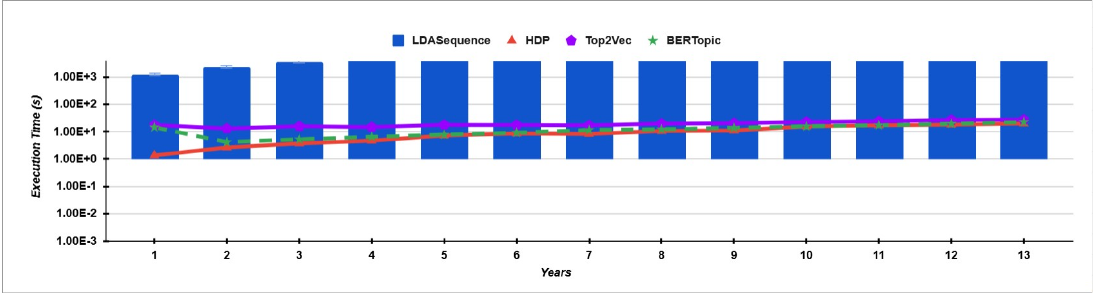}
    \caption{Computation time of models in the iterative process of updating the models. (a) location datasets (b) un-debate dataset }
    \label{fig:ExecutionTime}
\end{figure}

As can be seen in Figure \ref{fig:ExecutionTime}, HDP executes faster than others during scalability testing with location and un-debate datasets. 
HDP takes about $3$ minutes on average to complete an increment while Top2Vec and BERTopic take $4$ minutes and $9$ respectively. The parameters of HDP, Top2Vec, and BERTopic have no effect on the execution time. 
Conversely, the algorithm with the longest execution time for each increment is LDASequence, $6$ hours on average. 
This essentially indicates that to handle the enormous number of tuples, LDASequence requires additional resources and time. In addition, computation time of LDASequence is affected by number of topics specified.
This explains why, above, we did not evaluate LDASequence on the whole evaluation dataset (twitter and un-debate) because LDASequence was killed as it was running for days. 
For un-debate datasets, HDP shows the same behavior, except for computation time of Top2Vec being slightly higher when compared to BERTopic. 

\textit{}
\textit{}
\section{General Findings and Conclusion}
We presented a methodology to evaluate dynamic topic modeling algorithms namely LDASequence, HDP,  BERTopic, and Top2Vec on COVID-19 and UN-Debate dataset and compare them with regards to various metrics such as topic density, topic quality, topic evolution, topic stability, and execution time. Table \ref{table:general-findings} shows a summary of our evaluation.

\begin{table}[h]
  \centering
    \caption{Summary of evaluation results on the whole evaluation twitter/un-debate datasets.}
\resizebox{0.8\textwidth}{!}{
\begin{tabular}{llllllll} 
 \toprule
 Algorithm & \textit{Avg Coherence}& \textit{Avg Diversity}  & \textit{Avg Stability} & \textit{Avg. Execution Time}   \\ [0.5ex] 
 \midrule
 HDP & 0.30/0.36 & 0.15/0.38 & 1.0/1.0 & 58/20 mins/secs  \\
 BERTopic & 0.64/0.50 & 0.77/0.38  & 0.79/0.44 & 7/23 hours/secs   \\ 
 Top2Vec & 0.41/0.27 & 0.09/1.0 &  0.045/1.0& 1/36 hour/secs \\
 \bottomrule
\end{tabular}
\label{table:general-findings}}
\end{table}

\begin{table}[h]
  \centering
    \caption{Summary of evaluation results on the evaluation sub-dataset/un-debate.}
\resizebox{0.8\textwidth}{!}{
\begin{tabular}{llllllll} 
 \toprule
 Algorithm & \textit{Avg Coherence}& \textit{Avg Diversity} & \textit{Avg Stability} & \textit{Avg. Execution Time}   \\ [0.5ex] 
 \midrule
 LDASequence   & 0.31/0.40 & 0.67/0.76 & 0.98/0.96 & 6/3 hours \\
 HDP & 0.44/0.42 & 0.46/0.46 & 0.55/0.87 &  3/10 mins/secs \\
 BERTopic & 0.75/0.63 & 0.82/0.56 & 0.12/1.0   & 4/12 mins/secs   \\ 
 Top2Vec & 0.42/0.56 &  0.43/1.0 & 0.51/1.0 & 4/19 mins/secs \\
 \bottomrule
\end{tabular}
\label{table:general-findings}}
\end{table}

From our evaluation, we conclude that LDASequence is useful when the readability of extracted topics is the main goal (i.e., having control over the number of topics). 
However, LDASequence shows acceptable (but not high) topic quality.
As a challenge then, the number of topics has to be carefully selected.
The weakness of LDAsequence as observed is attributed to poor performance when the data scales up (i.e., with a huge number of tuples). Hence, in a limited hardware resource environment, LDASequence presents a high execution time. The execution time for HDP is minimal but it produces topics that are redundant as the number of tuples grows. The topics produced at every increment with a new month slice are entirely different or slightly related to the topics of other months. In a use case where execution delays are crucial and topic quality is tolerated, HDP is the option. BERTopic has a high topic quality 
because it automatically identifies the number of topics at every increment, it presents a large number of topics that are on the other hand difficult to read through. 
More efforts are needed to control the number of topics produced by carefully selecting hyper-parameters. 
Another drawback is that its computation time is relatively high. 
In a scenario where the topic quality and/or stability are/is important and the algorithm latency is tolerated, BERTopic is the best algorithm. Top2Vec presents better results in terms of topic evolution in comparison with other algorithms. 
Top2Vec also extracts topics automatically. In this case, it tends to produce more topics when the number of tuples grows. 
The topic quality is poor (the minimal compared to others) and it seems not to perform well in finding topics when datasets content are slightly similar in content. 
Also, another drawback of Top2Vec is observed during the initialization of the algorithm. 
For a small number of tuples, there is a need to double the number of tuples 
to initialize building of the model. 
In a usecase where the objective is to have high topic evolution in an incremental procedure, Top2Vec is a best choice, especially when other metrics are not considered important.

\bibliography{main}
\end{document}